\documentclass[preprint,showpacs,preprintnumbers,amsmath,amssymb]{revtex4}

\usepackage{graphicx}
\usepackage{dcolumn}
\usepackage{bm}

\begin{document}

\title{\large Conventional proximity effect in bilayers of\\
superconducting underdoped $La_{1.88}Sr_{0.12}CuO_4$ islands\\
coated with non superconducting overdoped
$La_{1.65}Sr_{0.35}CuO_4$}

\author{G. Koren}
\email{gkoren@physics.technion.ac.il} \affiliation{Physics
Department, Technion - Israel Institute of Technology Haifa,
32000, ISRAEL} \homepage{http://physics.technion.ac.il/~gkoren}

\author{O. Millo}
\email{milode@vms.huji.ac.il} \affiliation{Racah Institute of
Physics, The Hebrew University of Jerusalem, Jerusalem 91904,
Israel}

\date{\today}
\def\bfig {\begin{figure}[tbhp] \centering}
\def\efig {\end{figure}}

\normalsize \baselineskip=8mm  \vspace{15mm}

\begin{abstract}

Following a recent study by our group in which a large $T_c$
enhancement was reported in bilayers of the non-superconducting
$La_{1.65}Sr_{0.35}CuO_4$ and superconducting
$La_{1.88}Sr_{0.12}CuO_4$ films [Phys. Rev. Lett. \textbf{101},
057005 (2008)], we checked if a similar effect occurs when
superconducting $La_{1.88}Sr_{0.12}CuO_4$ islands are coated with
a continuous layer of the non superconducting
$La_{1.65}Sr_{0.35}CuO_4$. We found that no such phenomenon is
observed. The bare superconducting islands film behaves as an
insulator where transport occurs via hopping or tunneling between
islands, but exhibits a weak signature of localized
superconductivity at $\sim$18 K. When over coated with a
$La_{1.65}Sr_{0.35}CuO_4$ film, it becomes superconducting with a
$T_c$ onset of 16.6 K, which is less than that of a thick
$La_{1.88}Sr_{0.12}CuO_4$ film (25.7 K). We therefore conclude
that the lower $T_c$ in the bilayer is due to a conventional
proximity effect.

\end{abstract}


\maketitle

A recent paper by Yuli \textit{et al.} \cite{Yuli-1} has reported
the observation of a large $T_c$ enhancement in bilayers of the
non superconducting, heavily overdoped (OD)
$La_{1.65}Sr_{0.35}CuO_4$ (LSCO-35) and superconducting underdoped
(UD) $La_{1.88}Sr_{0.12}CuO_4$ (LSCO-12), in comparison with the
bare film of the latter. This study was based on the idea that in
the UD regime of the high temperature superconductors, $T_c$ is
determined by phase fluctuations, while pairing without phase
coherence occurs in the pseudogap regime at considerably higher
temperatures
\cite{Uemura,Kivelson,Corson,Ong-Nernst,ourreview,ARPES}, similar
to the case of granular superconductors \cite{Merchant}. In the OD
regime however, pairing and phase order occur simultaneously, with
a robust phase stiffness. Therefore, in the interface of bilayers
composed of UD and OD films, one can envision a scenario in which
the high phase stiffness of the OD layer locks via Josephson
coupling the phases of the preformed pairs in the UD layer. This
together with the high pairing in the UD layer is expected to lead
to a $T_c$ enhancement effect above that of both components
\cite{Kivelson2,Berg}. A large $T_c$ enhancements effect was
reported by Gozar et al. in a similar bilayer system of
$La_{1.55}Sr_{0.45}CuO_4$ and $La_2CuO_4$ \cite{Gozar}. It is
unclear to us whether this observation is due to the
aforementioned combination of UD  and OD layers, as it is
questionable whether the ozonized $La_2CuO_4$ which they used is
actually underdoped. Motivated by the these works, we turned to
study bilayers consisting of an insulating LSCO-12 islands film
(below the percolation threshold) coated by a continuous metallic
LSCO-35 layer. The fundamental question here is whether a similar
$T_c$ enhancement effect takes place also in such a granular
system. This question touches upon the wider issue of the
connection between nominally-granular and electronically-disorder
superconductors \cite{Deutscher}. In contrast to the case of
bilayers made of continuous LSCO-12 and LSCO-35 films [1], no
$T_c$ enhancement effect was found here when the LSCO-12 film was
granular. Although global phase-coherence was achieved upon the
deposition of LSCO-35 layer on top of the insulating LSCO-12
islands film and a resistive superconducting transition was
obtained, the corresponding transition temperature was smaller
than that of a continuous LSCO-12 film prepared under similar
conditions.  The transition temperature, however, was still lower
but close to the apparent $T_c$ of the localized superconductivity
in the islands ($\sim$18 K). Our results are thus in accord with a
conventional proximity effect where global phase coherence is
achieved via Josephson coupling between the localized
superconducting grains as discussed by Merchant et al.
\cite{Merchant}.\\

\begin{figure} \hspace{-20mm}
\includegraphics[height=9cm,width=14cm]{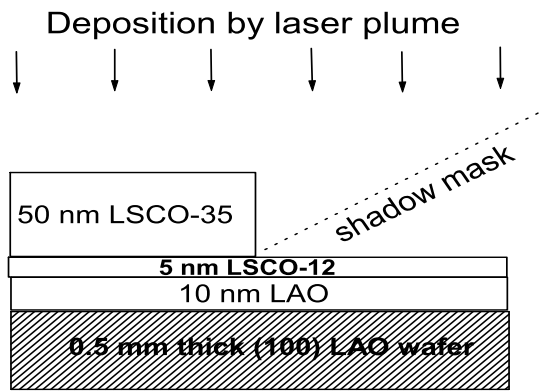}
\vspace{-0mm} \caption{\label{fig:epsart}The experimental setup
for \textit{in situ} deposition of a bilayer and its reference
film on the same wafer. }
\end{figure}

The bilayers and films in the present study were grown epitaxially
on (100) $LaAlO_3$ (LAO) wafers by standard laser ablation
deposition using the third harmonic of a Nd-YAG laser with 10 ns
laser pulse duration, 355 nm wavelength and 1.2 J/cm$^2$ laser
fluence on the cuprate targets. The optically polished LAO wafers
had $10\times 10$ mm$^2$ area and 0.5 mm thickness. Deposition was
done at 790 $^\circ$C wafer temperature while the ambient oxygen
pressure was maintained at 90 mTorr. Cooling was done in 0.75 atm
of oxygen pressure with a dwell of 2 hours at 450 $^\circ$C. In
order re-form the surface of our polished substrates and improve
their surface quality, we always deposited on them a thin
epitaxial LAO template layer of 10 nm thickness prior to the
deposition of the other layers. Then the samples were prepared
\textit{in-situ} as shown schematically in Fig. 1, where half of
the wafer is coated with the reference islands film (nominal 5 nm
LSCO-12), and the other half with the cap layer film (50 nm
LSCO-35) on top of the islands film, using a shadow mask.\\

\begin{figure}\hspace{-15mm}
\includegraphics[height=8cm,width=12cm]{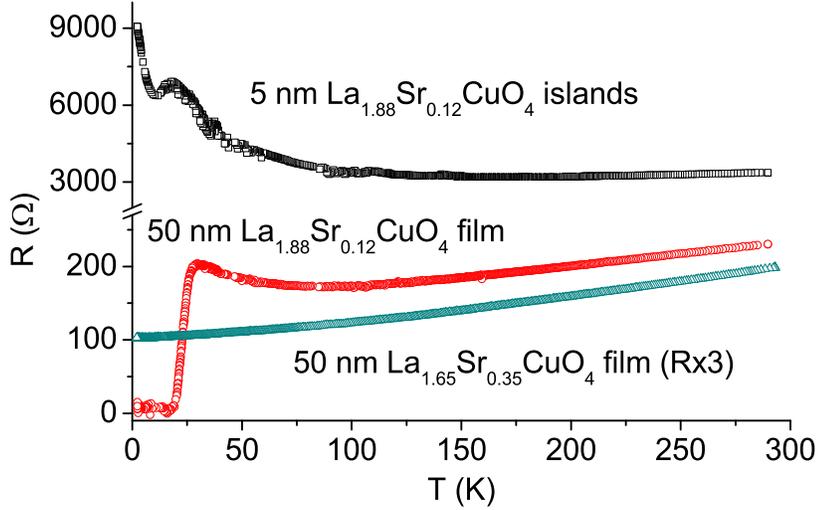}
\vspace{0mm}\caption{\label{fig:epsart}(Color online) Resistance
versus temperature of the three relevant reference films. The
nominally 5 nm thick LSCO-12 islands film, a 50 nm thick LSCO-12
film, and a 50 nm thick LSCO-35 film, all on 10 nm deposited LAO
layer on the (100) LAO wafers. }
\end{figure}

Since the nominal 5 nm thick LSCO-12 islands film is very
sensitive to humid air, its exposure to the atmosphere was kept to
a minimum (a few minutes), until it was mounted in the measuring
probe in He gas ambient. The resistance versus temperature (R(T))
was then measured using the standard 4-probe dc technique, and the
results are shown in Fig. 2. The resistance of this film shows a
metal to insulator transition at about 170 K, a signature of the
superconducting transition with a maximum at 18 K and minimum at
11 K, and an insulating behavior at lower temperatures. Such type
of 'reentrant superconductivity' behavior is typical of
superconductor island films, and is attributed to localized
superconductivity in the grains \cite{Goldman}. Although this
feature does not lend itself to an accurate determination of the
corresponding $T_c$, it appears to be significantly lower than
that of the continuous 50 nm thick LSCO-12 film (see Fig. 2). The
conductance of this film is therefore due to hopping and/or
tunneling between superconducting LSCO-12 grains. Hence we can
refer to this film as an "islands film", as opposed to a
continuously superconducting film (generally, when the film is
thicker), since no superconducting percolation path exists in it
between the contacts in the R(T) measurement. To further
characterize the surface morphology of this islands film, we show
in Fig. 3 an Atomic Force Microscope (AFM) image of a $2\times
2\,\mu m^2$ area. A typical height profile Z(nm) along the line
marked on the image is also shown. One can see that there are
holes in this film which are at least 3 nm deep, and possibly
deeper since the micro-lever of the AFM (the tip) is too wide to
penetrate them. The rougher area at the top left side of the
image, which has about twice the roughness of the smoother area,
appears as isolated islands of a few $\mu$m in size and occupy
only about 5\% of the total area of the film. We can thus conclude
from Figs. 2 and 3 that the connections between the grains or
islands in the nominally 5 nm thick LSCO-12 film are weak, and
that transport in this film via these weak links is controlled by
hopping or tunneling.\\

\begin{figure} \hspace{-15mm}
\includegraphics[height=8cm,width=12cm]{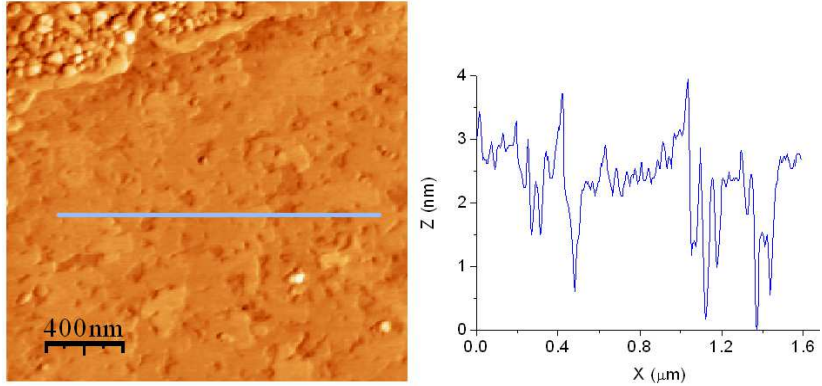}
\vspace{0mm}\caption{\label{fig:epsart}(Color online) An AFM image
of the nominally 5 nm thick LSCO-12 film, together with a typical
height profile along the line shown in the image. }
\end{figure}

\begin{figure}\hspace{-15mm}
\includegraphics[height=8cm,width=12cm]{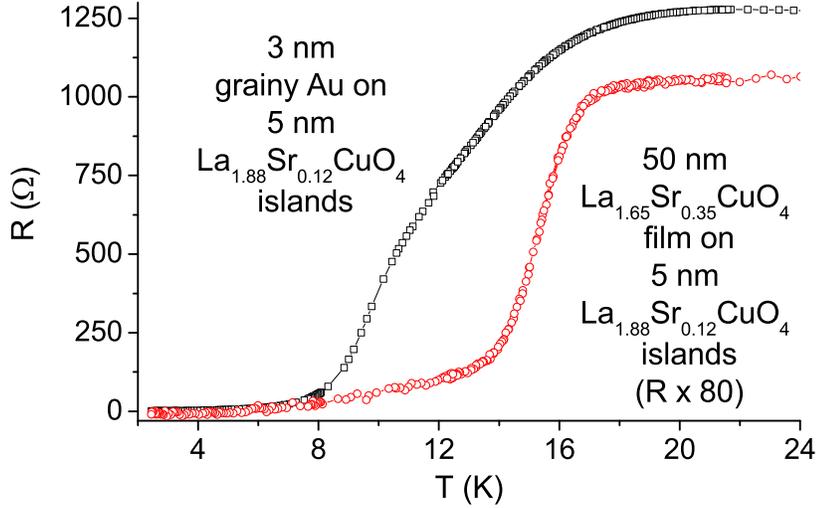}
\vspace{-5mm}\caption{\label{fig:epsart}(Color online) Resistance
versus temperature of the 3 nm Au/50 nm LSCO-35/5 nm LSCO-12
trilayer and the 3 nm Au/5 nm LSCO-12 bilayer. (The top 3 nm Au
layer in the trilayer is not marked in the figure).}
\end{figure}

We now turn to the main results of this study. After the nominally
5 nm thick LSCO-12 islands film was deposited on the whole wafer,
a 50 nm thick LSCO-35 film was deposited on half of the wafer by
the use of a shadow mask as shown in Fig. 1. Then the mask was
removed, and a nominally 3 nm thick layer of gold was deposited on
the whole wafer at the same deposition temperature (790 $^\circ$C)
and in 400 mTorr of oxygen gas pressure. This very thin gold
layer, which was composed of loosely connected ball-like grains,
was mainly used to improve the contacts in the transport
measurement, but also induced a proximity effect in the LSCO-12
islands film. The wafer was then patterned by wet acid etch to
separate its two halves by a $2\times 10\, mm^2$ wide strip along
the contact line of the mask with the wafer. The resistance versus
temperature of the two halves of this wafer was measured in the
same cooling run and is shown in Fig. 4. The onset of the
superconducting transition (defined here as the temperature of
10\% drop in resistance) of the Au/LSCO-12 bilayer and the
Au/LSCO-35/LSCO-12 trilayer are very similar, 15.6 and 16.6 K,
respectively. We note here that using the same definition of $T_c$
onset for the bare 5 nm thick LSCO-12 islands film of Fig. 2, one
also finds the same $T_c$ of 16.6 K for the loosely connected
grains. The respective mid-point transitions in Fig. 4 are already
quite different and equal to 11.6 and 15.3 K. The same behavior
and even more so is true for the extrapolated to zero-resistance
transition temperatures which are found at 8.3 and 13.7 K,
respectively. At lower temperatures, there is a Josephson 'tail'
or 'knee' of the resistance in both cases, which in the trilayer
reaches zero at 5-6 K, and in the bilayer decreases exponentially
and approaches zero at 2 K. The improved superconducting quality
of the trilayer compared to that of the bilayer is due to both the
thicker LSCO-35 film and the better Fermi wave-vector matching in
the former system. Our results indicate a clear proximity effect
and Josephson coupling in both cases, where, on the one hand
global phase-coherence was induced in the insulating 5 nm (base)
LSCO-12 islands film, but, on the other hand, $T_c$ was lower than
that of the bare 50 nm thick reference LSCO-12 film, as depicted
in Fig. 2. More importantly, $T_c$-onset in both bilayer and
trilayer films are comparable to the location of the 'reentrant'
peak feature observed in the R(T) curve of the islands film (Fig.
2), which is associated with the onset of localized
superconductivity in the LSCO-12 islands. Our data are similar and
in agreement with a previous study by Nesher \textit{et al.}
\cite{Nesher} where a proximity effect was observed in a bilayer
of $YBa_2Cu_3O_{7-\delta}$ islands over-coated with a normal
$YBa_2Cu_{2.7}Fe_{0.3}O_y$ layer (above its superconducting
transition temperature). It may be instructive to compare our
findings also with those of Merchant et al. \cite{Merchant}.
There, global phase-coherence was induced in an insulating Pb
islands film by over coating with a normal Ag layer. The $T_c$ of
this bilayer first increased with increasing Ag thickness up to
nearly the bulk $T_c$ of lead, but then decreased with further
\textit{in-situ} Ag deposition. The initial increase of $T_c$ was
attributed to enhanced intergrain Josephson coupling and the
subsequent decrease, to the Ag-Pb proximity effect.  Our results
are in partial agreement with these observations.  On one hand we
find emergence of global superconductivity upon over coating the
insulating LSCO-12 islands film by a metallic layer of either a
thin (3 nm) Au or a thick (50 nm) LSCO-35, in agreement with
\cite{Merchant}. On the other hand, the $T_c$-onset values in both
cases were similar ($\sim$16 K) and well below that of a bare
thick (50 nm) LSCO-12 film (25.7 K). It is possible that with some
intermediate thickness of the metallic ad-layer in the present
study, a $T_c$ close to that of the thick LSCO-12 film could be
achieved as found by Merchant \textit{et al.} for bulk Pb.
However, the fact that the $T_c$-onset in both our trilayer and
bilayer films were close to the reentrant superconductivity
feature of the LSCO-12 islands film, suggests that it is
determined by the $T_c$ of the grains.\\

\newpage

As we have seen in Fig. 2, the base LSCO-12 islands film has an
insulating behavior at low temperatures with no $T_c$, while the
ten times thicker and clearly continuous reference LSCO-12 film
has a $T_c$ onset of 25.7 K and $T_c(R=0)$ of 19.6 K. For the sake
of completeness, we also show in Fig. 2 the metallic but
non-superconducting behavior of the overdoped LSCO-35 film. Note
that due to the low resistance of this film, the resistance of the
curve shown is multiplied by a factor of 3. Another point to note
is that the normal resistance of the Au/LSCO-12 islands bilayer
just above the superconducting transition is about 80 times larger
than that of the Au/LSCO-35/LSCO-12 islands trilayer. This
indicates that the resistance of the nominally 3 nm thick Au layer
is much higher than that of the 50 nm LSCO-35 film, and
corroborate the islands nature of this layer too. The ball-like
(or islands) surface morphology of the Au layer was also seen in
several AFM images in this study and in other experiments where
gold films were deposited under similar conditions. We also note
that in the trilayer case, the top gold layer only reduces the
contact resistance and does not contribute to the proximity effect
which obviously occurs at the LSCO-12 and LSCO-35 interface,
located 50 nm beneath the Au grains.\\

To summarize our main finding, the transport results of Fig. 4 on
the bilayer and trilayer together with the results of the bare
reference layers of Fig. 2, demonstrate conventional proximity and
Josephson coupling effects between the LSCO-12 islands and either
the gold or the LSCO-35 layers, where the bilayer or trilayer
$T_c$ values are found in between those of the bare reference
films. Our findings here differ from our previous results, where a
$T_c$  enhancement with respect to that of a continuous LSCO-12
film, was found for both bilayers of 10 nm LSCO-12 on 90 nm
LSCO-35 and 10 nm LSCO-35 on 90 nm LSCO-12 \cite{Yuli-1}.  It
should be noted that in Ref. [1] all the layers were grown on
(100) $SrTiO_3$ (STO) wafers. In this case, thicker base layers
(either LSCO-35 or LSCO-12) were used in order to release tensile
strains with the STO substrate, so that the interface of the
bilayers will be less affected by strains. In the preset study, we
chose to work with an LAO substrate since its lattice match with
the various LSCO-x cuprates is better (cubic a axes are 0.3788 nm
for LAO versus 0.3905 nm of STO while LSCO-10 has a 0.3784 nm
lattice constant  \cite{Locquet}). The fact that no $T_c$
enhancement was found here, in contrast to Ref. [1], may be due to
the following reason. The enhanced $T_c$  in Ref. [1] was found to
be confined to a continuous two-dimensional interface layer
between the LSCO-12 and LSCO-35 films. In the present study, a
continuous two-dimensional interface layer cannot exist, since the
LSCO-12 islands film was below the percolation threshold.
Nevertheless, Josephson coupling between islands induced by either
the Au or LSCO-35 ad-layers was sufficient for the actual
emergence of global phase coherence in the bilayer and trilayer
films in the present study.\\

In conclusion, conventional proximity effect and
Josephson-coupling were observed in the present study in bilayers
of an insulating LSCO-12 islands film with either LSCO-35 or Au.
Although global phase-coherence developed in these films, the
transition temperatures appeared to be limited by the proximity
effect, showing no enhancement with respect to bare  LSCO-12
films.\\

{\em Acknowledgments:}  This research was supported in part by the
Israel Science Foundation (grant \# 1564/04), the Heinrich Hertz
Minerva Center for HTSC, the joint German-Israeli DIP project, the
Karl Stoll Chair in advanced materials, the Harry de Jur Chair in
applied science, and by the Fund for the Promotion of Research at
the Technion.\\

\bibliography{AndDepBib.bib}

\bibliography{apssamp}

\end{document}